%
%
\input harvmac
\input epsf
%

\let\\=\cr
\def\[{$$}
\def\]{$$}
\def\section{\newsec}
\def\subsection{\subsec}

\def\hline{\noalign{\hrule}}
\def\Ndash{--\nobreak}
\def\arXiv#1{\nopagenumbers\abstractfont\rightline{\tt #1}\vskip 1.2cm }
\def\title#1{\[\vbox{\baselineskip=20pt\halign{\titlefont%
     \hfill##\hfill\\#1\crcr}}\]\abstractfont \vskip .5cm \pageno=0}
\def\author#1{\centerline{\authorfont #1}}
\def\email{\footnote{$^\star$}{\tt zotov@eas.sinp.msu.ru}}
\def\address#1{\[\vbox{\baselineskip=14pt%
     \halign{\strut\hfill##\hfill\\#1\crcr}}\]%
     \bigskip\bigskip\bigskip}
\def\abstract#1{{\narrower\noindent#1 \par}\bigskip}

\def\today{\ifcase\month\or January\or February\or
   March\or April\or May\or June\or July\or August\or September\or
   October\or November\or December\fi
   \space\number\day, \number\year}



\def\date#1#2{{\narrower\noindent PACS: {#1}\hfill{#2}\par}\vfill%
\tenpoint\supereject\global\hsize=\hsbody%
\footline={\hss\tenrm\hyperdef\hypernoname{page}\folio\folio\hss}}

%
\def\caption#1#2{\medskip{\narrower\item{\bf Fig.\ #1:}{#2}\par}}
\def\figure#1#2#3{%
\centerline{\epsfysize=0.4\vsize \epsfbox{#1}}\caption{#2}{#3}}

\def\table#1#2{{\narrower\itemitem{\bf Table #1:}{#2}\par}\nobreak}

\def\footnoterule{\kern-3pt\hrule width 2.15truein \kern 2.6pt}
\def\frac#1#2{{#1\over#2}}

\def\sign{\mathop{\rm sgn}\nolimits}
\def\abs#1{\left| #1 \right|}
\def\const{\rm const}


\def\roughly#1{\mathrel{\raise.3ex\hbox{$#1$\kern-.75em\lower1ex%
\hbox{$\sim$}}}}


\def\jrnl#1#2#3#4#5#6{#1, #2, {\it #3\/} {\bf #4}, #5, #6.}
\def\jnx#1#2#3#4#5#6#7{#1, #2, {\it #3\/} {\bf #4}, #5, #6; {\tt #7}.}

\def\xxx#1#2#3{#1, #2, {\tt #3.}}


\lref\BMK{
 \jrnl{R. Bartnik and J. McKinnon}
 {Particle-like solutions of the Einstein--Yang--Mills equations}
 {Phys. Rev. Lett.}{61}{141\Ndash144}{1988}}

\lref\VGreview{
 \jnx{M. S. Volkov, D. V. Gal'tsov}{Gravitating non-Abelian
 solitons and black holes with Yang--Mills fields}{Physics Reports\/}
 {319}{1\Ndash83}{1999}{hep-th/9810070}}

\lref\wePRD{
 \jnx{E. E. Donets, D. V. Gal'tsov, M. Yu. Zotov}
 {Internal structure of Einstein--Yang--Mills black holes}{Phys. Rev.}
 {D56}{3459\Ndash3465}{1997}{gr-qc/9612067}}

\lref\VGLett{
 \jrnl{M. S. Volkov, D. V. Gal'tsov}
 {Non-Abelian Einstein--Yang--Mills black holes}
 {Pis'ma Zh. Eksp. Teor. Fiz.}{50}{312\Ndash315}{1989}
 [\it JETP Lett.\/ \bf 50, \rm 345 (1990)].}

\lref\KMuA{
 \jrnl{H. P. K\"unzle, A. K. M. Masood-ul-Alam}
 {Spherically symmetric static $SU(2)$ Einstein--Yang--Mills fields}
 {J.~Math. Phys.}{31}{928\Ndash935}{1990}}

\lref\Bizon{
 \jrnl{P. Bizo\'n}{Colored black holes}
 {Phys. Rev. Lett.}{64}{2844\Ndash2847}{1990}}

\lref\JMP{
 \jnx{M. Yu. Zotov}
 {Dynamical system analysis for the Einstein--Yang--Mills
 equations}{J.~Math. Phys.}{41}{4790\Ndash4807}{2000}
 {gr-qc/9906024}}

\lref\SWextend{
 \jnx{J. A. Smoller, A. G. Wasserman}
 {Investigation of the interior of colored black holes and the
 extendability of solutions of the Einstein--Yang/Mills equations
 defined in the far field}
 {Commun. Math. Phys.}{194}{707\Ndash732}{1998}{gr-qc/9706039}}

\lref\SWRNlike{
 \jnx{J. A. Smoller, A. G. Wasserman}
 {Reissner--Nordstr\"om--like solutions of the~$SU(2)$ Einstein--Yang/Mills
 equations}{J.~Math. Phys.}{38}{6522\Ndash6559}{1997}{gr-qc/9703062}}

\lref\CNS{
     \jnx{A. Corichi, U. Nucamendi, D. Sudarsky}
     {Einstein--Yang--Mills isolated horizons: phase space, mechanics,
     hair, and conjectures}{Phys. Rev.}{D62}{044046}{2000}{gr-qc/0002078}}

\lref\BHmdbh{
     \xxx{J. Bjoraker, Y. Hosotani}
     {Monopoles, dyons and black holes in the four-dimensional
     Einstein--Yang--Mills theory}{hep-th/0002098}}

\lref\Wasserman{
     \xxx{A. Wasserman}
     {Solutions of the spherically symmetric $SU(2)$
     Einstein--Yang--Mills equations defined in the far field}
     {gr-qc/0003046}}

\lref\OliKunzle{
 \xxx{T. A. Oliynyk, H. P. K\"unzle}{Local existence proofs for the
 boundary value problem for static spherically symmetric
 Einstein--Yang--Mills fields with compact gauge groups}{gr-qc/0008042}}

\lref\Mann{
 \jnx{R. B. Mann}
 {Black holes of negative mass}{Class. Quant. Grav.}{14}{2927\Ndash2930}
 {1997}{gr-qc/9705007}}

\lref\CSmass{
 \jnx{A. Corichi, D. Sudarsky}{Mass of colored black holes}{Phys. Rev.}
 {D61}{101501}{2000}{gr-qc/9912032}}

\lref\Octave{J. W. Eaton, {\it GNU Octave: A high-level interactive language
 for numerical computations},
 Edition~3 for version 2.0.13, 1997;  {\tt http://www.che.wisc.edu/octave/}}

\lref\NSzero{
 \xxx{U. Nucamendi, D. Sudarsky}{Black holes with zero mass}{gr-qc/0004068}}

\def\capOne{%
     The gauge field function for the solutions defined by the upper
     branch of~$w_3$, see Table~1.}
\nfig\RNupperw{\capOne}
\def\capTwo{%
     The gauge field function for the solutions defined by the lower
     branch of~$w_3$, see Table~2.}
\nfig\RNlowerw{\capTwo}
\def\capThree{%
     The effective charge function for the solutions shown in
     Figure~\xfig\RNupperw.  Numbers in the body of the plot denote
     the number of nodes of~$w$.}
\nfig\RNupperQ{\capThree}
\def\capFour{%
     The effective charge function for the solutions shown in
     Figure~\xfig\RNlowerw.  Numbers in the body of the plot denote
     the number of nodes of~$w$.}
\nfig\RNlowerQ{\capFour}
\def\capFive{%
     The dependence of $w_3$ and the ADM mass~$M$ on~$u_1$ near $M=0$
     for solutions with zero number of nodes of the gauge field function.
     Here $w_0 = -\frac{1}{2}$, and the upper branch of $w_3$ values
     is taken.}
\nfig\zeroADMa{\capFive}
\def\capSix{%
     The same as in Figure \xfig\zeroADMa, but $w_0 = -\frac{3}{2}$,
     and the lower branch of $w_3$ values is taken.}
\nfig\zeroADMb{\capSix}
\def\capSeven{%
     The gauge field function for the solutions with zero number
     of nodes.  ``Outer'' solutions have negative ADM mass, ``inner''
     solutions have positive ADM mass.
     Parameters of the solutions are given in Tables~1, 2, and~3.}
\nfig\FourWs{\capSeven}
\def\capEight{%
     The metric function $u$ for the solutions shown in Figure~\xfig\FourWs.
     Parameters of the solutions are given in Tables~1, 2, and~3.
     Dashed line: $y=r^2$.}
\nfig\FourUs{\capEight}

\arXiv{gr-qc/0011065}

\title{%
                    SOLUTIONS WITH NEGATIVE MASS \\
                    FOR THE $SU(2)$ EINSTEIN--YANG--MILLS\\
                    EQUATIONS
}

\author{%
                    Mikhail Yu.\ Zotov\email
}

\address{%
                    D. V. Skobeltsyn Institute of Nuclear Physics\\
                    Moscow State University, Moscow 119899, Russia
}

\abstract{%
     The aim of this note is to clarify the structure of nontrivial
     asymptotically flat solutions with nonpositive ADM mass for the
     static, spherically symmetric Einstein--Yang--Mills equations with
     $SU(2)$ gauge group.  The presented numerical results demonstrate
     that these solutions have zero number of nodes of the gauge field
     function.  This is a feature that is present neither for
     particlelike nor for black hole solutions in this model.
}

\date{04.25.Dm, 11.15.Kc}{December 1, 2000%
     \foot{Version 2: a statement concerning masses of black hole
     solutions on page~2 qualified, reference~12, and an acknowledgment
     added. Results unchanged.}}


\sequentialequations
\section{Introduction}

     Static, spherically symmetric Einstein--Yang--Mills (EYM) equations
     with gauge group $SU(2)$ attract considerable attention after the
     celebrated discovery of the particlelike solutions made by Bartnik
     and McKinnon in 1988~\BMK\ and the subsequent generalization of this
     result to the black hole case~\refs{\KMuA\VGLett\Ndash\Bizon}.
     A review of the results obtained in 1988--1998 and an extensive
     bibliography can be found in~\VGreview.  Some more recent results are
     presented in~\refs{\JMP\CSmass\CNS\BHmdbh\Wasserman\Ndash\OliKunzle}.
     But though a great amount of investigations of the EYM equations has
     already been performed, some questions still remain open.  To
     introduce the subject of this paper, recall some basic facts.

     First, the space-time metric for the static, spherically
     symmetric EYM equations can be written as
\[
     ds^2 = \sigma^2 N \, dt^2 - N^{-1} \, dr^2
            - r^2 \left(d \vartheta^2
                    + \sin^2 \vartheta \, d \varphi^2 \right),
\]
     where $N$ and~$\sigma$ depend on~$r$, and the Yang--Mills
     gauge field reads as
\[
     A = (T_2 \, d \vartheta - T_1 \sin \vartheta \, d \varphi) \, w
         + T_3 \cos \vartheta \, d \varphi ,
\]
     where $T_i = \frac{1}{2} \, \tau _i$ are the~$SU(2)$ group
     generators and $\tau _i$ are the Pauli matrices, $i = 1,2,3$
     (see, e.g.,~\VGreview).

     It is convenient to write down the EYM equations in this framework
     in the form of two ordinary differential equations for the metric
     function $u = r^2 N$ and the gauge field function~$w$:
\eqn\EYM{
     \eqalign{
          & r u'    - \bigl(1 - 2 {w'}^2 \bigr) u
                    + \left(1 - w^2 \right)^2 - r^2 = 0, \\
          & r u w'' - \left[u + \left(1 - w^2 \right)^2 - r^2 \right] w'
                    + \left(1 - w^2 \right) r w = 0,
     }
}
     and a decoupled equation for~$\sigma$:
\[
          r \sigma' - 2 \sigma {w'}^2 = 0.
\]
     Since \EYM\ do not involve~$\sigma$, one can use these to obtain~$u$
     and~$w$, and then solve the equation for~$\sigma$.  Thus the following
     analysis is restricted to Eqs.~\EYM.  Notice also that~\EYM\ are
     invariant under the transformation $r \to -r$; hence only the case
     $r \ge 0$ will be discussed.

     The mass function is usually defined as $m = (r-u/r)/2$; then the ADM
     mass is $M = \lim_{r\to\infty}m$.

     Now, recall that the EYM equations~\EYM\ admit two explicit
     solutions:  the Schwarzschild solution
\eqn\exactS{
     w \equiv \pm 1, \quad u = r^2 - 2 M r,
}
     and the Reissner--Nordstr\"om solution
\eqn\exactRN{
     w \equiv 0, \quad u = r^2 - 2 M r + 1,
}
     where $M$ is an arbitrary constant, representing the ADM mass of the
     corresponding solution.  In the first case, $m(r) \equiv M = \const$,
     and in the second one, $m = M - \frac{1}{2r}$.  Thus, both exact
     solutions admit arbitrary nonpositive values of the ADM mass.

     Next, both particlelike and black hole solutions of~\EYM\ are
     asymptotically flat and belong to the two-parameter family
\eqn\AsFlat{
     \eqalign{
     u & = r^2 - 2 M r + o(r), \\
     w & = w_{\infty} + b r^{-1} + o(r^{-1})
     }
}
     as $r \to \infty$, where $M$ and $b$ are arbitrary constants, and
     $w_\infty =\pm 1$.  More than
     this, it has recently been proved by Wasserman~\Wasserman\ that
     all real solutions of the EYM equations~\EYM\ defined at spatial
     infinity either have $w\equiv 0$ or belong to the family~\AsFlat.
     But while~\AsFlat\ includes solutions with nonpositive ADM mass,
     both particlelike and black hole solutions in this model have only
     strictly positive values of mass~\refs{\BMK\KMuA\VGLett\Ndash\Bizon}.
     Thus, there appears a natural question:  what are the nontrivial
     solutions of the family~\AsFlat\ that possess nonpositive ADM mass?
     This question is especially interesting in view of the results
     presented in~\refs{\Mann, \CSmass, \NSzero}.  (One may conjecture
     that these solutions are not globally defined, i.e., they exist only
     in the vicinity of $r=\infty$.  But this is not true in view of the
     result by Smoller and Wasserman, who have shown that all solutions
     defined in the far field are globally defined~\SWextend.)

     Finally, recall that for both particlelike and black hole solutions
     the gauge field function~$w$ reaches spatial infinity being within
     the strip $(-1, 1)$, and thus having $\sign b = \sign w_\infty$.
     This gives rise to another question: what are the solutions that
     have $\sign b = -\sign w_\infty$, so that $w$ reaches infinity
     staying outside the strip $(-1,1)$?  It is clear that in this case
     $w$ should have zero number of nodes, since it cannot have local
     maxima for $w>1$ and local minima for $w<-1$~\refs{\KMuA, \Bizon}.

     To find an answer to both questions, one should study
     Reissner--Nordstr\"om-like (RN-like) solutions, introduced by
     Smoller and Wasserman~\SWRNlike.  These solutions satisfy the
     following property: they are defined for all $r>0$, and
     $\lim _{r \to +0} (N,w,w') = (\infty, w_0, 0)$, $\abs{w_0} < \infty$.
     It is important that they may have zero number of nodes of the
     gauge field function~\SWRNlike.

     Notice that black hole solutions with the Reissner--Nordstr\"om
     interior structure, found numerically in~\wePRD, do also belong to
     the class of RN-like solutions.  Hence, the class of RN-like
     solutions is too rich for the purposes of this investigation.
     Thus, in what follows I consider only those RN-like solutions that
     have no horizons, i.e., the metric function~$u$ is positive for all
     $r>0$; let us call them ``horizonless'' for brevity.  In other
     words, only RN-like solutions that have naked singularity at
     $r=0$ will be discussed below.

\section{Main Results}

     In order to find an answer to the posed questions, one may perform
     numerical integration of~\EYM\ using expansion~\AsFlat\ at spatial
     infinity.  But it seems to be more interesting to start integration
     from the vicinity of the origin, $r=0$.  Recall that RN-like
     solutions belong to one of the following disjoint families of local
     solutions as $r\to0$~\refs{\SWRNlike, \JMP}:

     \item{1.} The two-parameter family of solutions with the Schwarzschild
     type singularity:
\eqn\Sas{
     \eqalign{
          u & = u_1 r + r^2 + o(r^2), \\
          w & = \pm 1 + w_2 r^2 + o(r^2),
     }
}
     where $u_1$ and~$w_2$ are arbitrary constants.  Solutions of
     this family correspond to the vacuum value of the Yang--Mills field
     $|w(0)|=1$ and generalize the Schwarzschild solution~\exactS.
     For the RN-like solutions, $u_1>0$.

     \item{2.} The three-parameter family of solutions with the
     Reissner--Nordstr\"om type singularity:
\eqn\RNas{
     \eqalign{
          u & = (1 - w_0^2)^2 + u_1 r + r^2 + o(r^2), \\
          w & = w_0  + \frac{w_0}{2 (1 - w_0^2)} \, r^2 + w_3 r^3 + o(r^3),
     }
}
     where $w_0$, $u_1$, and $w_3$ are arbitrary constants, $w_0\ne\pm1$.
     These solutions correspond to the metric of the mass~$m_0 = - u_1/2$
     and the (magnetic) charge $Q_0^2 = (1 - w_0^2)^2$.  Obviously, they
     generalize the Reissner--Nordstr\"om solution~\exactRN.

\medskip

     First, consider solutions that belong to the family~\RNas.
     It is convenient to divide the set of all possible initial data
     $(w_0, u_1, w_3)$, $\abs{w_0} \ne 1$, into two subsets:
     $\abs{w_0}<1$ and $\abs{w_0}>1$.  Let us begin with the first one.

     Take $w_0 = -\frac{1}{2}$ and $u_1 = 1$ for convenience.
     For this set of initial parameters one can find two discrete
     families of horizonless RN-like solutions
     with $n=0,1,2,3,\dots$ nodes of the gauge field function, so that
     there are two branches of~$w_3$ that give rise to these solutions,
     see Tables~1, 2 and Figures~\xfig\RNupperw, \xfig\RNlowerw.
     Notice that one of the nodeless solutions possesses negative ADM
     mass, namely, a solution defined by~$w_3$ from the upper branch.

     Notice also that for the family shown in Figure~\xfig\RNlowerw\
     (Table~2) solutions with $n>0$ have an extra local minimum in
     comparison with the similar solutions from another family.  This,
     in particular, results in a more complicated behavior of the effective
     charge function $Q^2 = 2 r (M - m)$, cf.\ Figures~\xfig\RNupperQ\
     and~\xfig\RNlowerQ.

     A situation similar to the above can be observed for other~$w_0$,
     $\abs{w_0}<1$.  Namely, horizonless RN-like solutions do always
     come in pairs, and every time when there exists a solution with
     negative mass, it is unique [for fixed $(w_0,u_1)$] and
     necessarily has zero number of nodes.  The corresponding~$w_3$
     always belongs to the upper branch of its values.  (As a result, no
     solutions with negative mass have been found for $w_0 = 0$.)
     Surprisingly enough, but not a single solution with negative ADM
     mass has been found for $n>0$.

     A natural question that appears at this point is the following:
     is there a continuous transition from negative to positive values
     of the ADM mass?  To answer the question one has just to study the
     dependence of mass on~$u_1$. The answer appears to be affirmative,
     see Figure~\xfig\zeroADMa.

     To study the case $\abs{w_0}>1$, take $w_0 = -\frac{3}{2}$ and
     $u_1=1$ for symmetry.  In this case, one finds only two pairs of
     horizonless RN-like solutions, each pair containing solutions with
     $n=0$ and $n=1$, see Table~3.  Similar to the above, one of the
     solutions with zero number of nodes of the gauge field function has
     negative ADM mass.  But contrary to the above, $w_3$ that
     gives rise to this solution lies on the lower branch of its values.
     And again, studying the dependence of mass on~$u_1$ one can find
     continuous transition from solutions with negative mass to solutions
     with positive mass, see Figure~\xfig\zeroADMb.

     It is interesting to compare nodeless solutions with negative ADM
     mass with their counterparts that have positive ADM mass.  Though
     in both cases the gauge field function is monotonic, its behavior
     considerably differs in these two cases.  Namely, solutions with
     positive ADM mass get close to the asymptotic value much faster
     than their counterparts with negative mass as $r\to\infty$, see
     Figure~\xfig\FourWs.  It is also remarkable that the sign of mass
     of the corresponding solution seems to be completely determined by
     the behavior of the metric function~$u$ near the origin.  As one
     can see from Figure~\xfig\FourUs, $u$ is monotonic and lies above
     the parabola $y=r^2$ for both solutions with negative ADM mass.
     In contrast with this, for the solutions with positive mass, $u$
     has a local minimum not far from $r=0$, deep enough to put~$u$ below
     the curve $y=r^2$.  (A similar behavior of~$u$ was found for
     solutions with $n>0$.)  Compare also the effective charge functions
     for $n=0$ in Figures~\xfig\RNupperQ\ and~\xfig\RNlowerQ.

     Notice that solutions which start from $w_0 = -\frac{3}{2}$ have
     $\sign b = - \sign w_\infty$ (cf.\ \AsFlat).  Thus, the presented
     solutions provide an answer to both questions stated above.

\table{1}{%
     Parameters of horizonless RN-like solutions found
     for $w_0 = -\frac{1}{2}$, ${u_1 = 1}$; the upper branch of~$w_3$
     is given.
}
\[
\vbox{
     \offinterlineskip
     \halign{
            \strut \vrule \hfil\quad$#$\quad\hfil &
                   \vrule \hfil\quad$#$\quad\hfil &
                   \vrule \hfill\quad$#$&$#$\quad\hfill \vrule \\
            \hline
            n               &     w_3      &&  \hphantom{0} M  \\
            \hline
            0               & 0.354 748 1  & -0.&676 226  \\
            \hline
            1               & 2.356 651 7  &  0.&620 136  \\
            \hline
            2               & 2.964 024 6  &  0.&928 290  \\
            \hline
            3               & 3.109 680 9  &  0.&987 930  \\
            \hline
            }
     }
\]
\smallskip
\table{2}{%
     Parameters of horizonless RN-like solutions found
     for $w_0 = -\frac{1}{2}$, $u_1 = 1$; the lower branch of~$w_3$
     is given.
}
\[
\vbox{
     \offinterlineskip
     \halign{
            \strut \vrule \hfil\quad$#$\quad\hfil &
                   \vrule \hfil\quad$#$\quad\hfil &
                   \vrule \hfil\quad$#$\quad\hfil \vrule  \\
            \hline
            n               &     w_3       &    M        \\
            \hline
            0               & -8.384 981 5  &  0.190 093  \\
            \hline
            1               & -9.497 540 2  &  0.854 186  \\
            \hline
            2               & -9.786 684 1  &  0.975 527  \\
            \hline
            3               & -9.838 216 3  &  0.995 995  \\
            \hline
            }
     }
\]
\smallskip
\table{3}{%
     Parameters of horizonless RN-like solutions found
     for $w_0 = -\frac{3}{2}$, $u_1 = 1$.
}
\[
\vbox{
     \offinterlineskip
     \halign{
            \strut \vrule \hfil\quad$#$\quad\hfil &
                   \vrule \hfil\quad$#$&$#$\quad\hfil &
                   \vrule \hfill\quad$#$&$#$\quad\hfill\vrule \\
            \hline
                n     && \hphantom{0} w_3  && \hphantom{0} M  \\
            \hline
                0     &-0.&697 576 2  & -1.&578 504  \\
            \hline
                1     & 1.&280 047 0  &  0.&846 658  \\
            \hline
                1     & 5.&438 372 7  &  0.&860 686  \\
            \hline
                0     & 6.&969 363 8  &  0.&182 617  \\
            \hline
            }
     }
\]
\bigskip

     It is necessary to say a few words about the dependence of the
     structure of horizonless RN-like solutions on the value of~$u_1$.
     Begin with the case $w_0 = -\frac{1}{2}$.  As~$u_1$ becomes negative,
     then soon after the ADM mass becomes positive, horizonless
     solutions begin to disappear.  First, there disappear solutions
     with $n=0$.  
     Next comes the order of $n=1$ solutions, etc.
     Finally, besides singular solutions, there remain only solutions
     with a regular horizon (for $u_1\roughly<-1.8$).  A qualitatively
     similar situation takes place for other values of~$w_0$,
     $\abs{w_0}<1$.

     The situation with $w_0 = -\frac{3}{2}$ is in a sense opposite to the
     above.  As we have seen, for $u_1=1$ there exist only solutions with
     $n=0,1$.  But the number of horizonless RN-like solutions increases
     as~$u_1$ grows.  First, there appears a pair of $n=2$ solutions, then
     a pair of $n=3$ solutions, etc., and finally one finds two families of
     solutions each having an arbitrary number of nodes of the gauge field
     function.  Conversely, the horizonless solutions disappear as~$u_1$
     becomes smaller.  First, there disappear solutions with $n=1$ and then
     with $n=0$.  A qualitatively similar picture can be found in a wide
     range of~$w_0$, $\abs{w_0}>1$.  In all cases, only solutions with
     $n=0$ possess nonpositive mass.

     For solutions with negative ADM mass, the value of~$M$ decreases
     unboundedly as~$u_1$ grows, e.g, for the case $w_0=-\frac{3}{2}$ the
     dependence is nearly linear and reads as follows:
     $M \approx -u_1/2 + \const$.


     In fact, the dependence of the structure of RN-like solutions and
     their ADM masses on the values of $(w_0,u_1)$ is interesting enough
     to be the subject of a separate investigation, but goes far beyond
     the scope of the present note ({\it vita brevis, fisica longa}).

     It is necessary to stress that for all values of~$w_1$ and~$u_1$
     considered above one can find not only horizonless (and singular)
     solutions, but also solutions with a regular horizon.
     They are not discussed here since they are either singular (i.e,
     cannot be continued towards $r=\infty$) or correspond to black hole
     solutions.

     Finally, consider solutions with the Schwarzschild type
     singularity~\Sas.  In this case, one can also find horizonless RN-like
     solutions.  The numerical results presented above demonstrate that
     only solutions with zero number of nodes of~$w$ can have nonpositive
     values of the ADM mass.  But nontrivial solutions that satisfy
     expansion~\Sas\ at $r=0$ cannot have $n=0$.  Hence, it seems quite
     natural that solutions with nonpositive ADM mass and the Schwarzschild
     type singularity were not found.

     The performed numerical investigation reveals that a situation
     qualitatively similar to the discussed above can be found for other
     initial values~$w_0$ and~$u_1$.  This allows us to put forward the
     following conjecture: {\it For each fixed pair
     $(w_0,u_1)$ in\/} \RNas, {\it there exists at most one global
     horizonless solution with nonpositive ADM mass;
     each such solution has zero number of nodes of the gauge field
     function.}

\bigskip\centerline{{\bf Acknowledgments}}\nobreak

     I thank D.~V.~Gal'tsov for a useful discussion.
     I also thank Piotr Bizo\'n for kindly sending me one of his articles
     on the EYM equations when this investigation was in progress
     and Alejandro Corichi for useful communication.

     I gratefully acknowledge a very useful discussion with Robert Mann,
     who have kindly pointed me out an ambiguity in the statement
     concerning masses of black hole solutions in the first version
     of this paper.

     The work was supported in part by Russian Foundation for Basic
     Research grant No.\ 00-02-16306.


\pageinsert
\figure{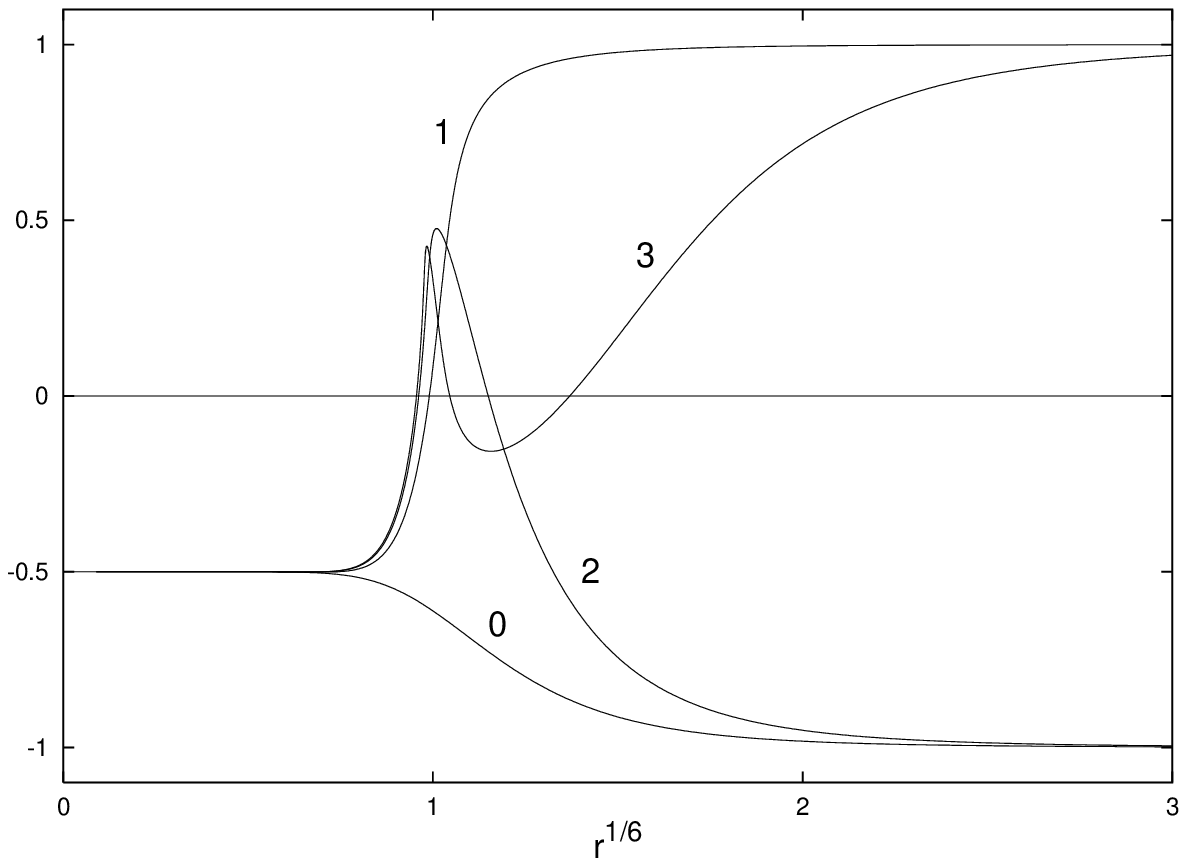}{\xfig\RNupperw}{\capOne}
\bigskip
\figure{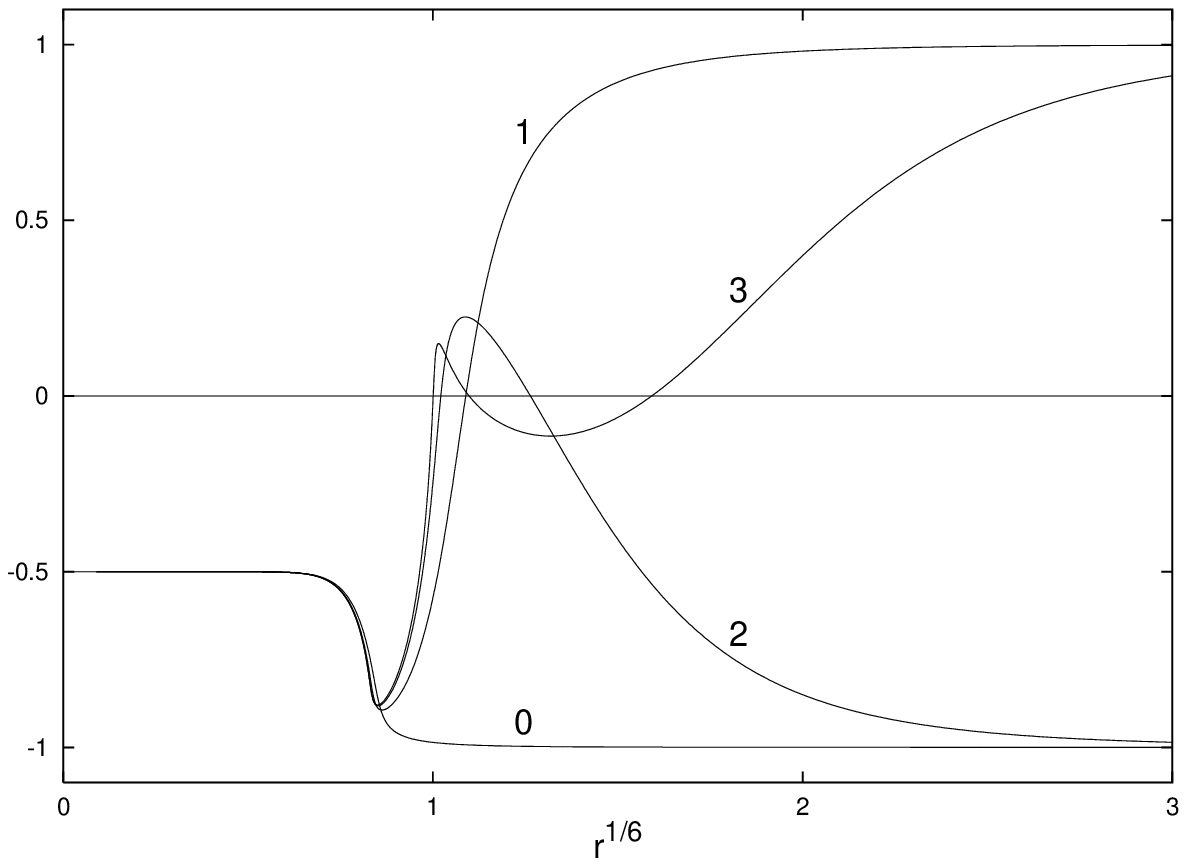}{\xfig\RNlowerw}{\capTwo}
\endinsert
\pageinsert
\figure{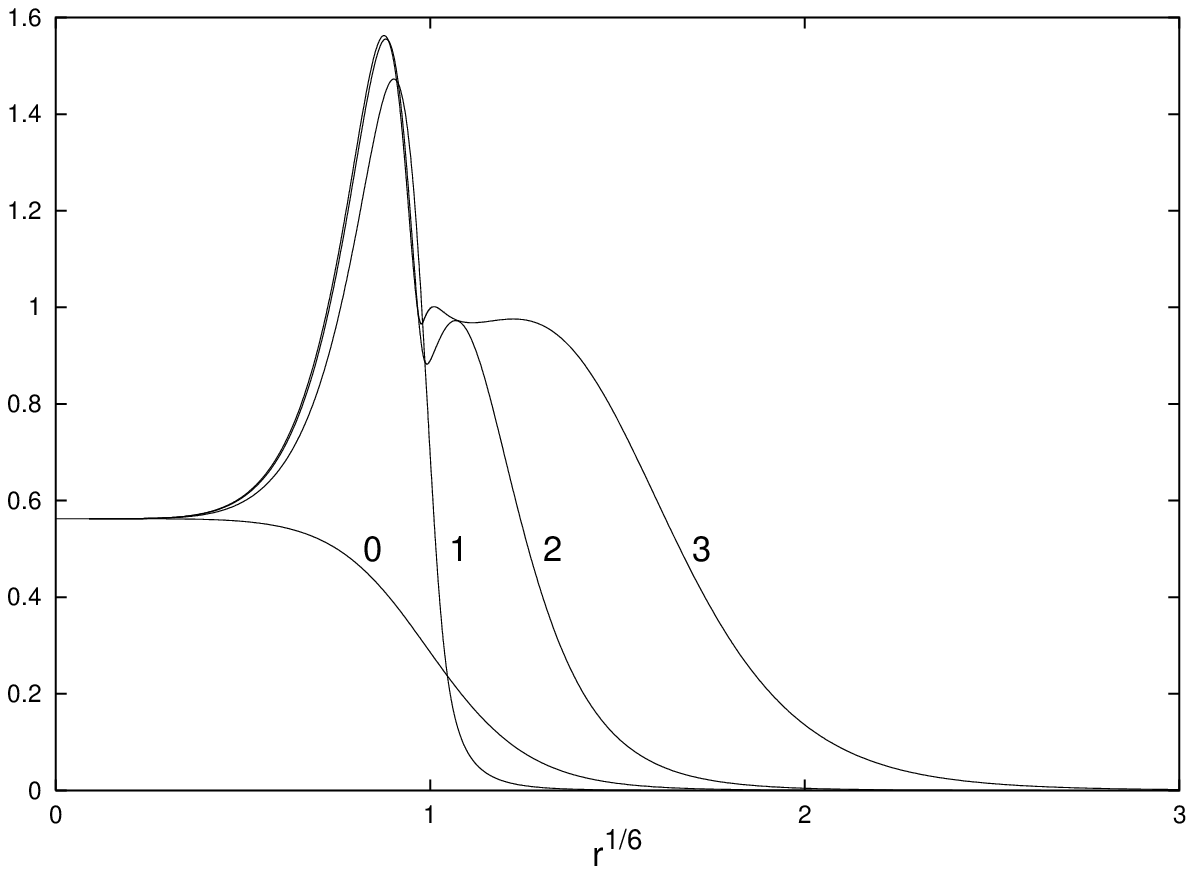}{\xfig\RNupperQ}{\capThree}
\bigskip
\figure{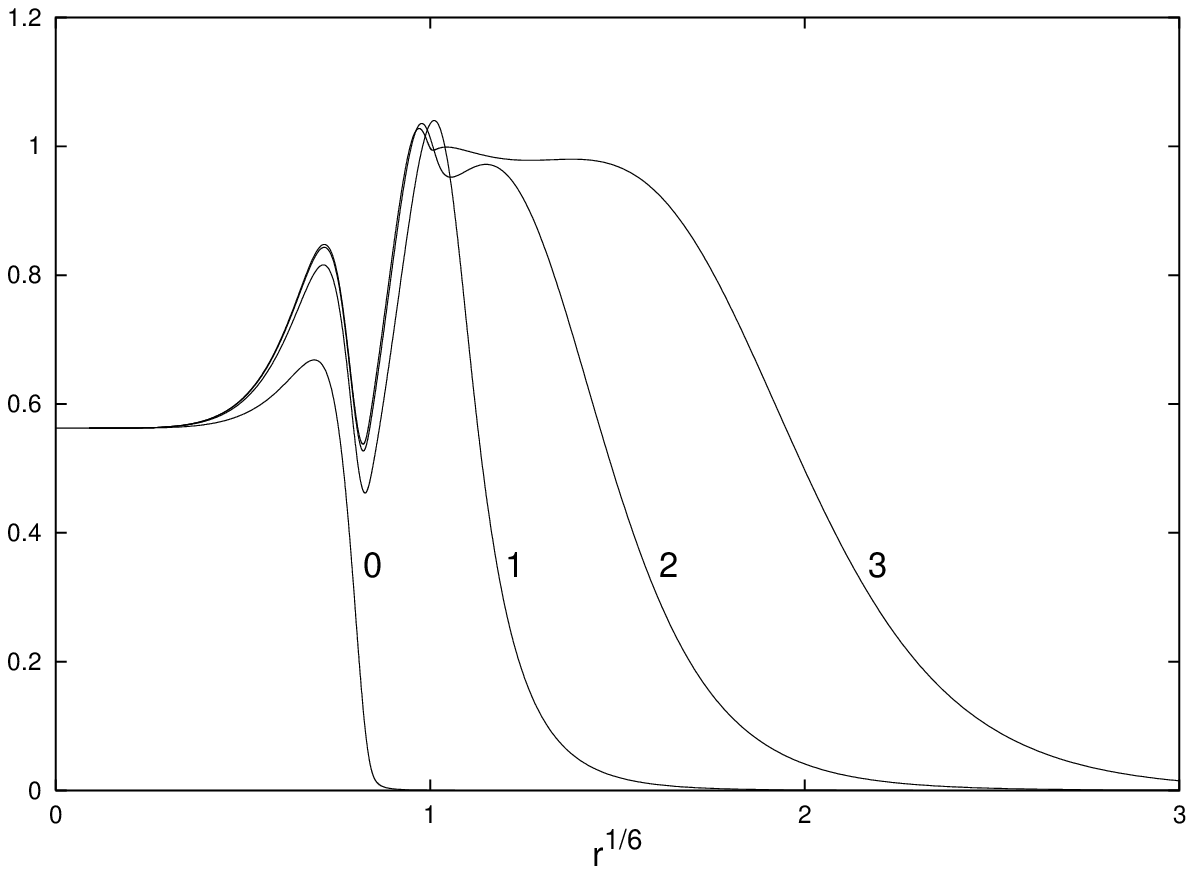}{\xfig\RNlowerQ}{\capFour}
\endinsert
\pageinsert
\figure{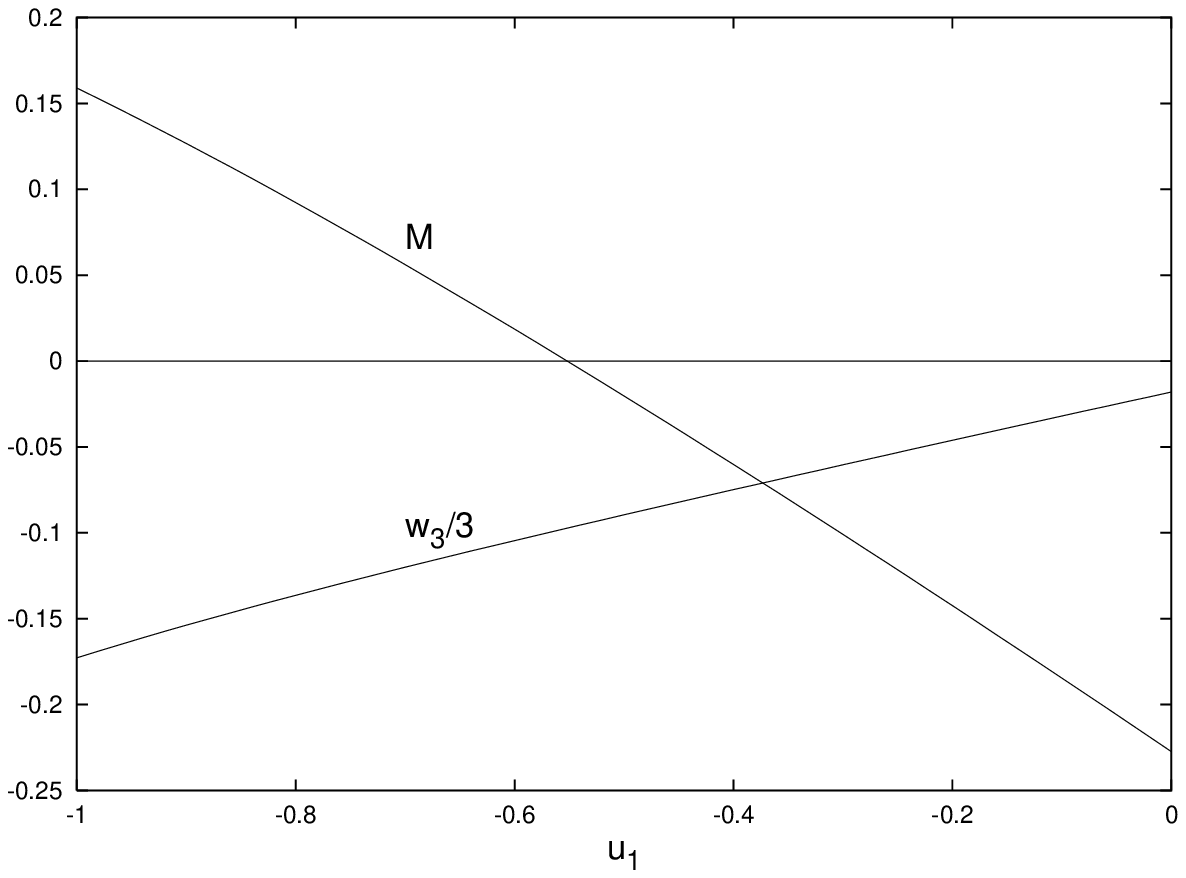}{\xfig\zeroADMa}{\capFive}
\bigskip
\figure{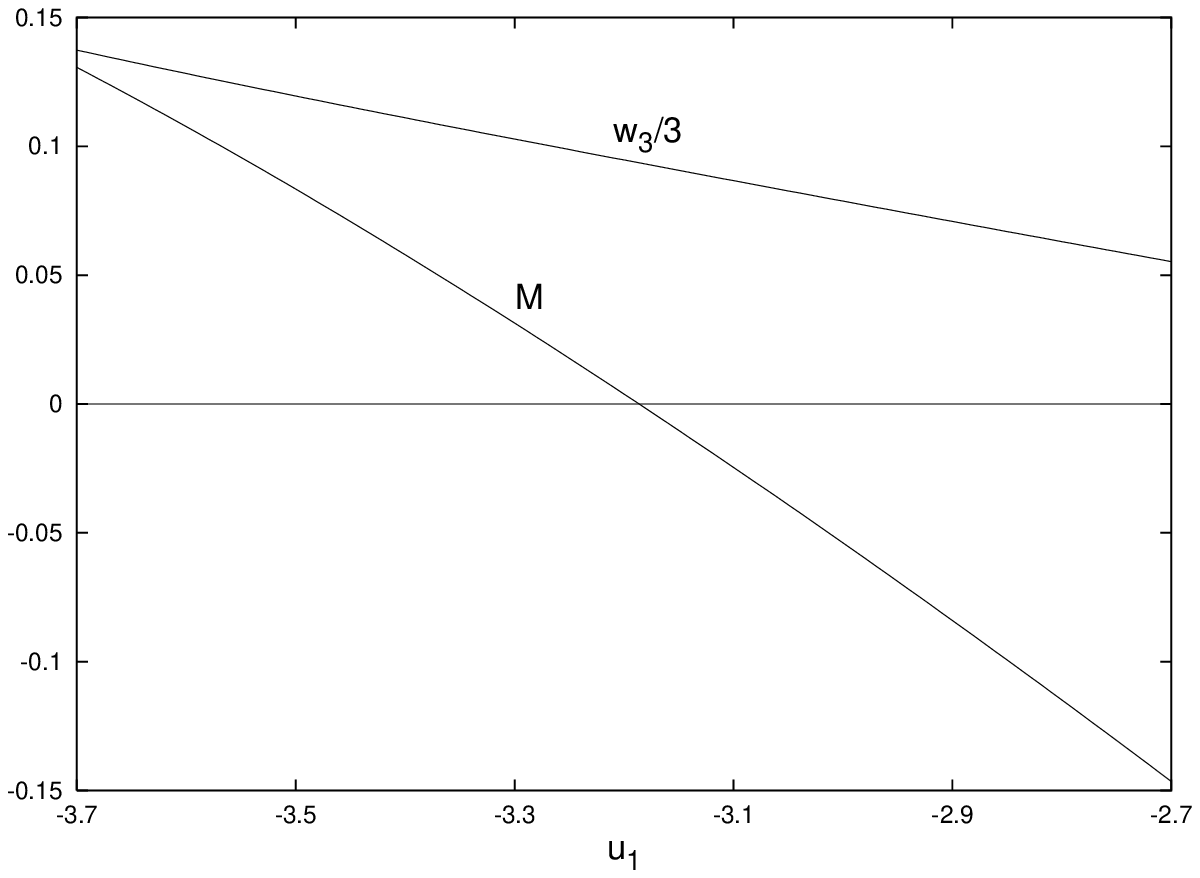}{\xfig\zeroADMb}{\capSix}
\endinsert
\pageinsert
\figure{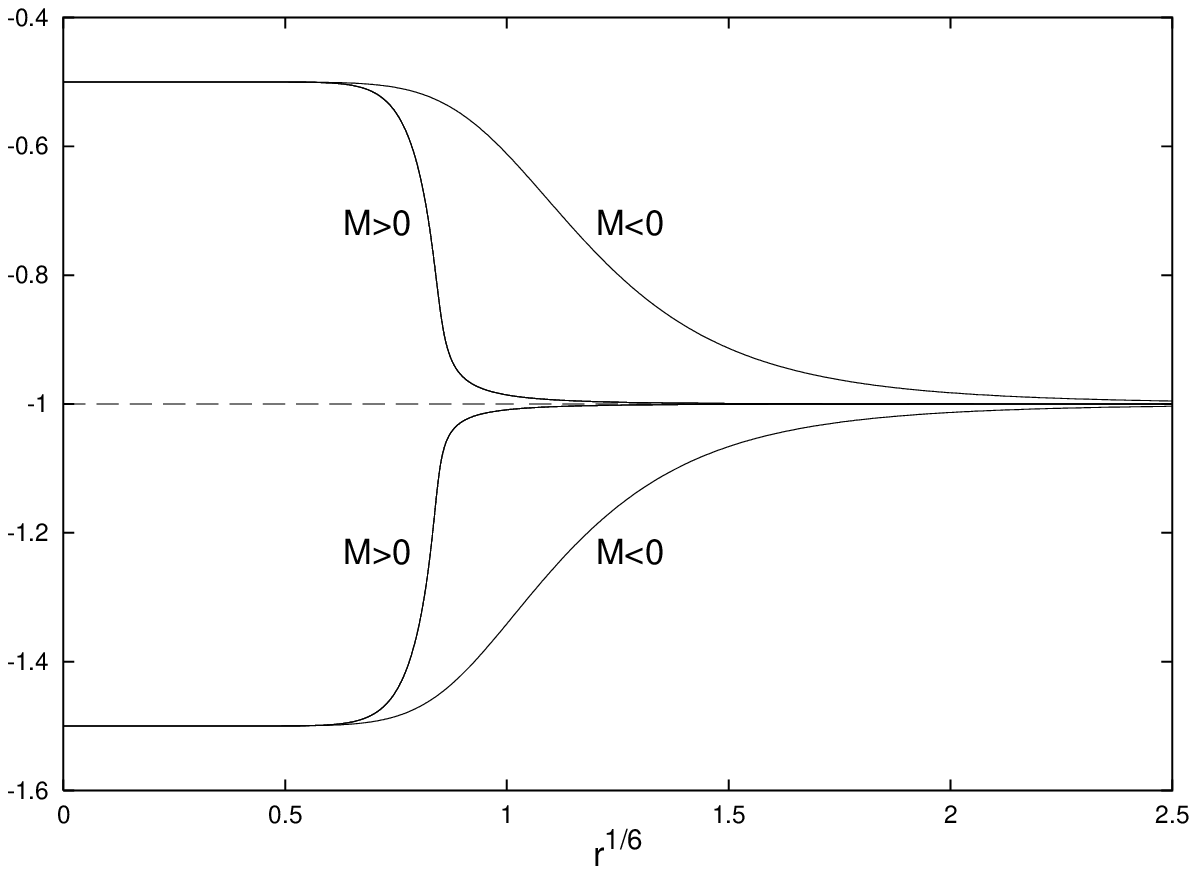}{\xfig\FourWs}{\capSeven}
\bigskip
\figure{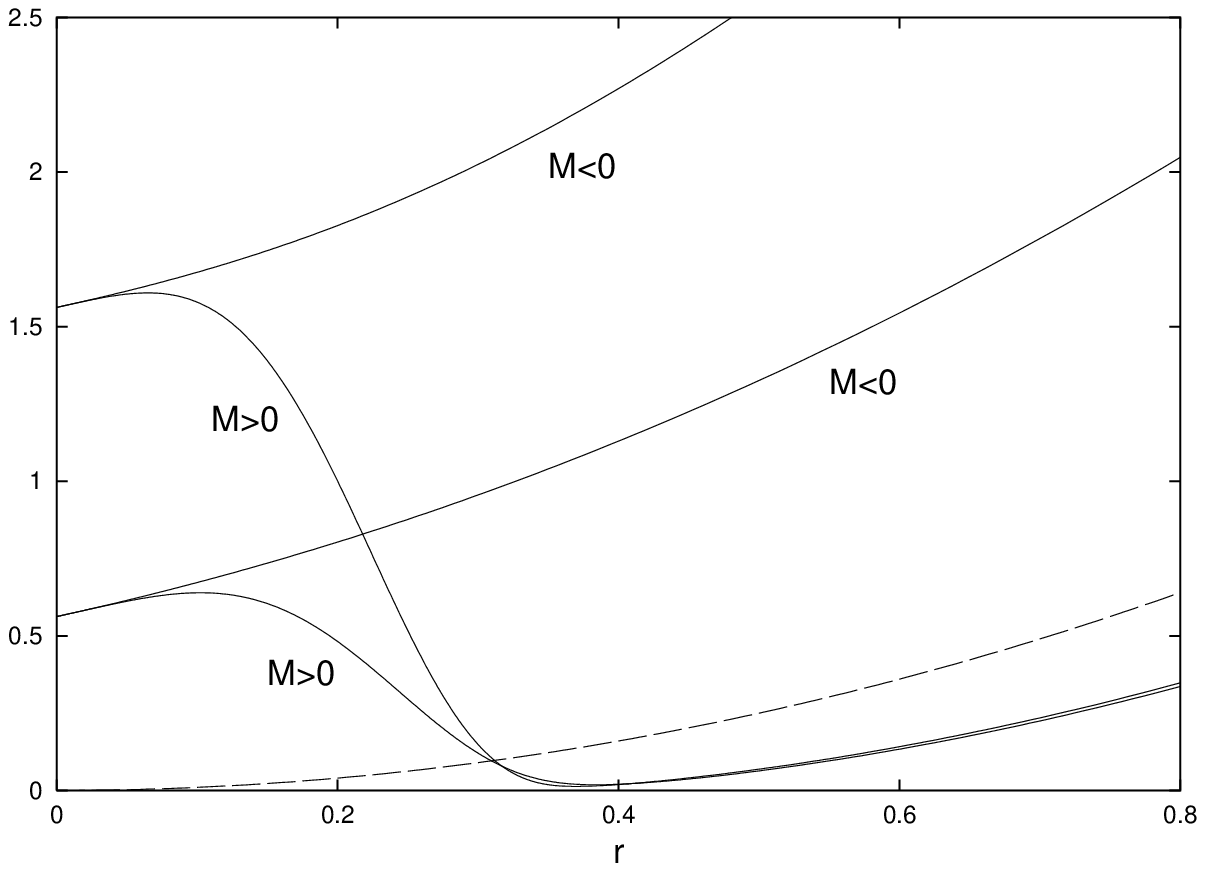}{\xfig\FourUs}{\capEight}
\endinsert
\listrefs
\bye